\documentclass[twocolumn,aps,prb,widetext,showpacs,superscriptaddress]{revtex4}
\usepackage{mathrsfs,bm,multirow,amsmath,amsfonts,amssymb,array,booktabs,float}
\usepackage{graphicx,times,graphics,color,epsfig}
\begin{document}
\title{Electronic structures of III-V zinc-blende semiconductors from atomistic first principles}
\author{Yin Wang}
\email{yinwang@hku.hk}
\affiliation{Department of Physics and the Center of Theoretical and Computational Physics, The University of Hong Kong, Pokfulam Road, Hong Kong SAR, China}
\author{Haitao Yin}
\email{wlyht@126.com}
\affiliation{Department of Physics and the Center of Theoretical and Computational Physics, The University of Hong Kong, Pokfulam Road, Hong Kong SAR, China}
\affiliation{Key Laboratory for Photonic and Electronic Bandgap Materials of Ministry of Education, School of Physics and Electronic Engineering, Harbin Normal University, Harbin 150025, China}
\author{Ronggen Cao}
\email{caoronggen@fudan.edu.cn}
\affiliation{Department of Physics and the Center of Theoretical and Computational Physics, The University of Hong Kong, Pokfulam Road, Hong Kong SAR, China}
\affiliation{Department of Materials Science, Fudan University, Shanghai 200433, China}
\author{Ferdows Zahid}
\affiliation{Department of Physics and the Center of Theoretical and Computational Physics, The University of Hong Kong, Pokfulam Road, Hong Kong SAR, China}
\author{Yu Zhu}
\affiliation{Nanoacademic Technologies, Brossard, PQ, Canada, J4Z 1A7}
\author{Lei Liu}
\affiliation{Nanoacademic Technologies, Brossard, PQ, Canada, J4Z 1A7}
\author{Jian Wang}
\affiliation{Department of Physics and the Center of Theoretical and Computational Physics, The University of Hong Kong, Pokfulam Road, Hong Kong SAR, China}
\author{Hong Guo}
\affiliation{Center for the Physics of Materials and Department of Physics, McGill University, Montreal, PQ, Canada, H3A 2T8}
\affiliation{Department of Physics and the Center of Theoretical and Computational Physics, The University of Hong Kong, Pokfulam Road, Hong Kong SAR, China}
\date{\today}
\begin{abstract}
For analyzing quantum transport in semiconductor devices, accurate electronic structures are critical for quantitative predictions. Here we report theoretical analysis of electronic structures of all III-V zinc-blende semiconductor compounds. Our calculations are from density functional theory with the semi-local exchange proposed recently [F. Tran and P. Blaha, Phys. Rev. Lett. {\bf 102}, 226401 (2009)], within the linear muffin tin orbital scheme. The calculated band gaps and effective masses are compared to experimental data and good quantitative agreement is obtained. Using the theoretical scheme presented here, quantum transport in nanostructures of III-V compounds can be confidently predicted.
\end{abstract}
\maketitle

\section{introduction}

 During the past five decades, semiconductor device miniaturization has brought modern device technology to the nanometer scale where quantum phenomena of charge and spin transport start to dominate the device physics.\cite{ITRS, Datta95} For tiny devices, the atomic nature of the materials are playing an increasingly prominent role\cite{ITRS} and charge transport in these systems driven by external fields is intrinsically a nonequilibrium problem. As a result the operation of nano-scale electronics crucially depends on the close coupling of nonequilibrium quantum transport phenomena with the atomic structure of the device material. Such a situation poses serious challenges to theoretical understanding and computational modeling of nanoelectronic device physics. More recently, atomistic methods have been combined with the Keldysh nonequilibrium Green's function (NEGF) formalism to meet the challenges of quantitative analysis of nanoelectronics. The atomistic methods are used to determine the material properties as well as the device Hamiltonian while NEGF is used to predict the nonequilibrium density matrix and transport properties. When combined self-consistently, such techniques can predict not only qualitative but also quantitative properties of quantum transport in nano-devices. Along this line, the state-of-the-art formalism is to carry out self-consistent density functional theory (DFT) atomistic calculation within the NEGF framework.\cite{jeremy} So far, parameter-free NEGF-DFT methods have been widely applied to capture quantum transport physics from the atomistic point of view.\cite{DattaBook}

 However, in order to apply first principles approach to investigate \emph{semiconductor} nanoelectronics, some very serious issues have to be resolved. First, realistic semiconductor devices (e.g. transistors) have large number of atoms and doped with small concentrations of impurities, while typical DFT methods can only comfortably deal with low hundreds of atoms. Second, DFT with the local density approximation (LDA)\cite{LDA, LDACA, Vosko80} and generalized gradient approximation (GGA) \cite{Perdew92b, GGAPBE96} underestimate band gaps of semiconductors. One could not predict transport results if band gaps and dispersions were not accurate: this is especially serious for semiconductors appearing in transistors since their gaps are not very large to begin with. Advanced methods such as GW\cite{GW} and hybrid functional\cite{HSE} can yield accurate band gaps for many systems, but require very large computation for semiconductor devices having hundreds to thousands of atoms.

For pure semiconductors, a recently proposed modified Becke-Johnson (MBJ) semilocal exchange was shown to give good band gaps for many semiconductors with a computational cost similar to that of LDA.\cite{MBJ} MBJ is not a fundamental solution to the issue of electron correlation, but it is practically useful for calculating band structures thus helpful for analyzing quantum transport properties. By implementing the MBJ functional into a newer generation of NEGF-DFT technique which is based on linear muffin tin orbital (LMTO) method with the atomic sphere approximation (ASA),\cite{nanodsim} transport in Si nano-transistors with localized doping and large number of atoms was recently analyzed.\cite{jesse-2012} Clearly, a very important next step is to investigate III-V semiconductors.

The III-V compound semiconductors are the most important materials for optoelectronic device applications and are also very important for the complementary metal-oxide-semiconductors (CMOS) technology. Major efforts are devoted by the microelectronics industry to integrate III-V semiconductors into Si CMOS.\cite{book} III-V semiconductors have been extensively investigated both experimentally and theoretically,\cite{Vurgaftman, Vurgaftman2, Franciosi} with particular attention paid to their band topologies since band parameters are critical for understanding quantum transport. Electronic structures of several direct gap III-V compounds have also been calculated using the MBJ functional as implemented in the planewave DFT codes. \cite{Kim} It is the purpose of this work to employ the DFT-MBJ approach within the LMTO-ASA scheme, to accurately calculate the band parameters of \emph{all} the zinc-blend III-V semiconductors. Our calculated band gaps at high-symmetry points ($\Gamma$, $X$, and $L$) are quantitatively compared with the corresponding experimental data; our calculated effective hole/electron masses at $\Gamma$ point are also well consistent to the experimental values. Details of the ASA schemes and the MBJ potentials will also be presented.

The rest of the paper is organized as follows. In the next section the calculation method is briefly discussed. Section III presents the results and Section IV is a short summary.

\section{Method}

Our calculation is based on the DFT-MBJ self-consistent approach where DFT is within the LMTO scheme and the atomic sphere approximation,\cite{LMTO} as implemented in the Nanodsim software package.\cite{nanodsim} For technical details of the Nanodsim algorithm we refer interested readers to the original literature.\cite{youqi, nanodsim}  A $12\times12\times12$ $k$-mesh were used to sample the Brillouin Zone of the primitive cell. The lattice constant for the semiconductors were adopted from Ref.~\onlinecite{Vurgaftman} and listed in Table~\ref{tab1}. In order to carry out LMTO DFT calculations for semiconductors, a good ASA scheme is very helpful. In our ASA, vacancy spheres were placed at appropriate locations as done in Ref.~\onlinecite{GaN} for space filling, and the same sphere radius are used for all the vacancy spheres and atomic spheres. Electrons in the full d-orbital of Ga, In, As, and Sb are included as valence electrons. After the LMTO DFT self-consistent calculation is completed, band structures are calculated by the the muffin-tin orbital (MTO) approach. The effective hole mass $m_h^*$ and effective electron mass $m_e^*$ were then obtained by fitting the valence band maximum and conduction band minimum to a parabola, respectively. Spin-orbit coupling was not considered in this work.

Some details of the MBJ semilocal exchange potential is worth mentioning. The MBJ potential has the following form,\cite{MBJ}
\begin{equation}\label{eq-MBJ}
v_{x,\sigma}^{MBJ}(r)=cv_{x,\sigma}^{BR}(r)+(3c-2)\frac{1}{\pi}\sqrt{\frac{5}{12}}
\sqrt{\frac{2t_{\sigma}(r)}{\rho_{\sigma} (r)}},
\end{equation}
where subscript $\sigma$ is spin index, $\rho_{\sigma}$ is the electron density, $t_{\sigma}$ is the kinetic energy density, and $v_{x,\sigma}^{BR}(r)$ is the Becke-Roussel potential.\cite{BR} The relative weight of the two terms is given by a parameter $c$ which depends linearly on the square root of $|\nabla \rho|/ \rho$. For all solids investigated by the MBJ potential so far,\cite{MBJ} it appears that $E_g$ increases monotonically with $c$. The $c$-value can be determined by the protocol discussed in Refs.~\onlinecite{MBJ, IMBJ}. In our calculations, because the LMTO-ASA is a site oriented technique, it allows the $c$-value to be ``local", namely one can use a $c$-value for the atom and another $c$-value for the vacancy sphere. In particular, for a given compound, the same $c$-value is used for the real atoms (e.g. Ga and As), and another $c$-value is used for the vacancy spheres. Recently, this MBJ scheme was applied in Ref.~\onlinecite{bandoffset} to determine very accurate band gaps for the Al$_x$Ga$_{1-x}$As compounds and band offsets for the GaAs/Al$_x$Ga$_{1-x}$As heterojunctions for the entire range of the concentration $0\leq x\leq 1$. Here we use the same scheme for all the III-V compounds and the optimized $c$-values are found and listed in Table~\ref{tab1}.
Even though one may expect more accurate band parameters by using different optimized $c$-values for all the different atomic spheres, the scheme we use is a good compromise between being simple and also reasonably accurate.

\begin{table}[t]
\caption{The lattice constant $\alpha$ of the III-V compounds and the $c$-values used in the  DFT-MBJ calculations. $c_{atom}$  and $c_{vac}$ are the $c$-values for real atoms and the vacancy spheres, respectively.}
\centering
\begin{tabular} {cccccccc}
\hline\hline
Material    &$\alpha({\AA})$    &$c_{atom}$ &$c_{vac}$  &Material    &$\alpha({\AA})$    &$c_{atom}$ &$c_{vac}$  \\ \hline
GaAs        &5.6533             &1.20       &1.39       &GaP         &5.4505             &1.13       &1.50       \\
AlAs        &5.6611             &1.11       &1.44       &AlP         &5.4672             &0.69       &1.44       \\
InAs        &6.0583             &1.18       &1.00       &InP         &5.8697             &1.04       &1.39       \\
GaSb        &6.0959             &1.17       &1.23       &GaN         &4.50               &1.19       &1.50       \\
AlSb        &6.1355             &1.12       &1.33       &AlN         &4.38               &1.59       &1.55       \\
InSb        &6.4794             &1.19       &0.62       &InN         &4.98               &1.56       &1.39       \\
\hline\hline
\end{tabular}
\label{tab1}
\end{table}

\section{Results}
\begin{table}[t]
\caption{Energies of the conduction band minima (band gaps, $E_g$) at the $\Gamma$, $X$, and $L$ points with respect to the valence band maximum at the $\Gamma$ point in units of eV, calculated by the DFT-MBJ approach at zero temperature. The experimental results are taken from Ref.~\onlinecite{Vurgaftman} (except where noted). MBJ$_V$ shows the DFT-MBJ values from Ref.~\onlinecite{Kim}. The minus sign of the deviation indicates that the calculated value is smaller then the experimental value.}
\centering
\begin{tabular} {cccccc}
\hline\hline
Material    &$E_g$    &This work  &MBJ$_V\cite{Kim}$    &Expt.\cite{Vurgaftman}    &Deviation(\%)  \\ \hline
GaAs        &$\Gamma$ &1.529      &1.52            &1.519    &0.7            \\
            &$X$      &2.003      &2.00            &1.981    &1.1  \\
            &$L$      &1.682      &1.72            &1.815    &-7.3  \\
AlAs        &$\Gamma$ &3.087      &                &3.099    &-0.4  \\
            &$X$      &2.240      &                &2.240    &0  \\
            &$L$      &2.800      &                &2.460    &13.8  \\
InAs        &$\Gamma$ &0.416      &0.43            &0.417    &-0.2  \\
            &$X$      &1.440      &2.01            &1.433    &0.5  \\
            &$L$      &1.225      &1.43            &1.133    &8.1  \\
GaP         &$\Gamma$ &2.887      &                &2.886    &0  \\
            &$X$      &2.350      &                &2.350    &0  \\
            &$L$      &2.429      &                &2.720    &-10.7  \\
AlP         &$\Gamma$ &3.635      &                &3.630    &0.1  \\
            &$X$      &2.513      &                &2.520    &-0.3  \\
            &$L$      &3.030      &                &3.570    &-15.1  \\
InP         &$\Gamma$ &1.421      &1.42            &1.424    &-0.2  \\
            &$X$      &2.502      &2.34            &2.384    &4.9  \\
            &$L$      &2.007      &2.11            &2.014    &-0.3  \\
GaSb        &$\Gamma$ &0.818      &0.82            &0.812    &0.7  \\
            &$X$      &1.117      &1.21            &1.141    &-2.1  \\
            &$L$      &0.876      &0.87            &0.875    &0.1  \\
AlSb        &$\Gamma$ &2.346      &                &2.386    &-1.7  \\
            &$X$      &1.698      &                &1.696    &0.1  \\
            &$L$      &1.845      &                &2.329    &-20.8  \\
InSb        &$\Gamma$ &0.238      &0.25            &0.235    &1.3  \\
            &$X$      &0.613      &1.52            &0.630    &-2.7  \\
            &$L$      &0.477      &0.82            &0.930    &-48.7  \\
GaN         &$\Gamma$ &3.298      &                &3.299    &0  \\
            &$X$      &4.528      &                &4.520    &0.2  \\
            &$L$      &5.997      &                &5.590    &7.3  \\
AlN         &$\Gamma$ &5.853      &                &6.0      &-2.5 \\
            &$X$      &4.908      &                &4.9      &0.2  \\
            &$L$      &9.304      &                &9.3      &0  \\
InN         &$\Gamma$ &0.781      &                &0.78\cite{Vurgaftman2}  &0.1\\
            &$X$      &3.456      &                &2.51     &37.7  \\
            &$L$      &4.635      &                &5.82     &-20.4  \\
\hline\hline
\end{tabular}
\label{tab2}
\end{table}

\begin{table}[t]
\caption{Effective hole and electron masses at the $\Gamma$ point in units of the electron rest mass $m_e$ calculated along the [100] direction. The experimental values are in the parentheses and taken from Ref.~\onlinecite{Vurgaftman} (except where noted). The minus sign of the deviation indicates that the calculated value is smaller than the experimental value.}
\centering
\begin{tabular} {ccccc}
\hline\hline
Material    &$|m^*_h|$      &Deviation(\%)  &$|m^*_e|$      &Deviation(\%)\\   \hline
GaAs        &0.355(0.350)   &1.4            &0.076(0.067)   &13.4\\
AlAs        &0.510(0.472)   &8.1            &0.149(0.150)   &-0.7\\
InAs        &0.373(0.333)   &12.0           &0.029(0.026)   &11.5\\
GaSb        &0.265(0.250)   &6.0            &0.041(0.039)   &5.1 \\
AlSb        &0.367(0.357)   &2.8            &0.118(0.140)   &-15.7\\
InSb        &0.263(0.263)   &0              &0.018(0.014)   &28.6\\
GaP         &0.411(0.326)   &26.1           &0.163(0.130)   &25.4\\
AlP         &0.595(0.518)   &14.9           &0.209(0.220)   &-5.0\\
InP         &0.450(0.532)   &-15.4          &0.094(0.080)   &17.5\\
GaN         &0.927(0.855)   &8.4            &0.232(0.15)    &54.7\\
AlN         &1.587(1.020)   &55.6           &0.319(0.25)    &27.6\\
InN         &0.977(0.833)   &17.3           &0.085(0.07\cite{Vurgaftman2}) &21.4\\

\hline\hline
\end{tabular}
\label{tab3}
\end{table}

\begin{figure}
\includegraphics[width=\columnwidth]{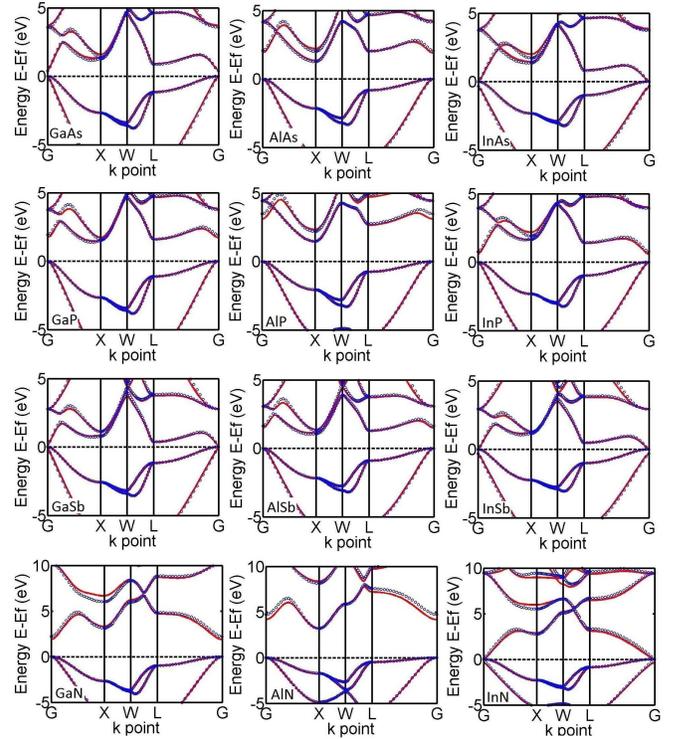}\\
\caption{(color online) The band structures for the III-V compound semiconductors obtained with LDA.  Red line is obtained by VASP, blue circles obtained by Nanodsim.}
\label{fig1}
\end{figure}

We begin by calculating the band structures of the III-V compounds with LDA\cite{LDA} using the projector augmented waves (PAW) method and a plane-wave basis set of 400 eV, as implemented in the Vienna \emph{ab initio} simulation package (VASP);\cite{VASP} as well as using the LMTO-ASA method discussed in the last Section where ASA is given in Ref.~\onlinecite{GaN}, as implemented in the Nanodsim\cite{nanodsim} NEGF-DFT package. Results for all the zinc-blende III-V compounds are plotted in Fig.~(\ref{fig1}), showing a perfect agreement of the valence bands and a very good agreement of the conduction bands between these methods, confirming that our ASA scheme in the LMTO-ASA  is accurate for calculating physical properties of these materials. It should be noted from Fig.~(\ref{fig1}) that the band gaps were underestimated by LDA.

\begin{figure}
\includegraphics[width=\columnwidth]{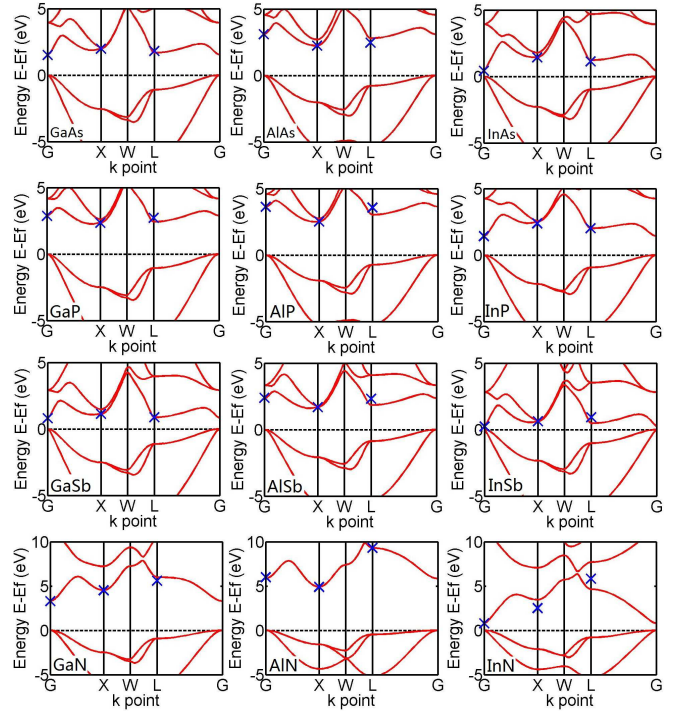}\\
\caption{(color online) The band structures for the III-V compound semiconductors obtained with MBJ.  Blue crosses indicate the experimental values.}
\label{fig2}
\end{figure}

Having confirmed our ASA scheme, we next apply the MBJ functional\cite{MBJ, IMBJ} to calculate the electronic structure again using the LMTO-ASA approach, and the MBJ results are plotted in Fig.~(\ref{fig2}). Compared with the LDA bands in Fig.~(\ref{fig1}), the opening of band gaps is evident. From Fig.~(\ref{fig2}), the band gap values at $\Gamma$, $X$ and $L$ points are obtained and listed in the third column of Table~\ref{tab2}. In Ref.~\onlinecite{Kim}, five direct gap compounds were calculated by the MBJ functional within a planewave method, their band gap values are also listed in Table~\ref{tab2} as the fourth column. Our results for these five compounds agree very well with the planewave results except for the $X$ point of InAs and InSb. As discussed in Ref.~\onlinecite{Kim}, for these narrow band semiconductors, it is difficult to accurately determine the gaps at the $X$ point. For instance, experimental gap value of $0.63$~eV was recommended in Ref.~\onlinecite{Vurgaftman} but other values such as $1.80$~eV was also reported in earlier literature\cite{drube}.

The experimental gap values are taken from Ref.~\onlinecite{Vurgaftman, Vurgaftman2} and listed as the fifth column in Table~\ref{tab2}. The last column is the percentage difference between our calculated values and the experimental values. Of the 36 tabulated gap values, 32 of them are in very good agreement with the experimental data. There are 4 values having a difference over 20\%: the $L$ point of InSb; the $X$ and $L$ points of InN; and the $L$ point of AlSb. These differences could be due to the uncertainties of the experimental values for the narrow gap compounds InSb and InN\cite{Kim} and, of course, the approximative nature of the DFT calculations. Nevertheless, it is impressive that, all in all, the MBJ band gaps are in very good consistency to the experimental values for these wide range of materials.

From the calculated band structures shown in Fig.~(\ref{fig2}), we obtain the effective masses of electrons and holes by fitting the valance band maximum or conduction band minimum to a parabola at the $\Gamma$ point along the [100] direction, the results are listed in Table~\ref{tab3} together with the experimental values.\cite{Vurgaftman, Vurgaftman2} The deviation of our calculated values to the experimental values are also listed for each compound. Very good agreement is obtained for most situations except for AlN and the electron effective mass of GaN. Given the fact that effective masses are usually very hard to determine accurately, the good consistency to the experimental results for most situations is very satisfactory.

\section{summary}

We have calculated the electronic structures of all zinc-blende III-V semiconductor compounds from density functional theory with the semi-local exchange of the MBJ form,\cite{MBJ} using the LMTO-ASA scheme. In our method and due to ASA, vacancy spheres are added to fill the volume of the semiconductors. Since this is a sited oriented calculation method, the weight parameter $c$ in the MBJ potential for the atoms and for the vacancy sites are different. We determine the optimal values of this weight parameter for all the compounds. The calculated band gaps are mostly in very good agreement with the corresponding experimental data. The obtained effective masses are also largely in good agreement to the measurements. For analyzing quantum transport in semiconductor nanoelectronics, accurate electronic structures are very important for quantitative predictions. In addition, as we have shown recently, the band-offset of heterojunctoins can also be accurately predicted using the same method as reported here.\cite{bandoffset} Together with the results of this paper, quantum transport properties of III-V systems can now be confidently calculated from the atomic point of view since the LMTO-ASA method can self-consistently calculate very large number of atoms.\cite{jesse-2012}

This work is supported by the University Grant Council (Contract No. AoE/P-04/08) of the Government of HKSAR, NSERC (HG), IRAP (YZ, LL) of Canada, and Reserve Talents of Universities Overseas Research Program of Heilongjiang (HY). We thank CLUMEQ for the computation resources.

\end{document}